\documentclass[12pt]{iopart}
\usepackage{graphicx}
\usepackage{hyperref}
\usepackage{amsfonts}
\usepackage{subfigure}
\usepackage{bm}
\input epsf

\begin{document}

\title{Vacuum Faraday effect for electrons}

\author{Colin Greenshields, Robert L. Stamps, Sonja Franke-Arnold} \ead{s.franke-arnold@physics.gla.ac.uk}
%\email[e-mail: ]{c.greenshields.1@research.gla.ac.uk}
\address{SUPA School of Physics and Astronomy, University of Glasgow, Glasgow G12 8QQ, UK}

\date{\today}

\begin{abstract} 
The optical Faraday effect describes the rotation of linear polarization 
upon propagation through a medium in the presence of a longitudinal 
magnetic field. The effect arises from a different phase delay 
between the right and left handed polarisation components of the light. 
In this paper we predict a Faraday effect for a completely different system: 
electron vortices.  Free electron vortex states were recently observed in transmission electron microscopy 
experiments, and they introduce new degrees of freedom into the probing of matter with electron 
beams. We associate a rotation of a vortex superposition with 
the fact that different phases are acquired by oppositely handed 
vortices propagating in a magnetic field.  We show that, in contrast to the optical 
Faraday effect, the rotation of the electron beam occurs in vacuum and 
arises from the intrinsic chirality of the constituent vortex 
states.
\end{abstract}


\maketitle

\section{The optical Faraday effect and its generalisation for electron waves}

Michael Faraday reported in 1845 that the polarisation of light can be affected by magnetic fields, an effect that now bears his name.  Since then,
the Faraday effect has found numerous metrological and research applications, including the ultra-sensitive detection of magnetic fields,
\cite{PhysRevA.47.1220, PhysRevA.64.033402}, or of fields generated by electron plasmas in interstellar space and the ionosphere \cite{warwick1964far, Mendillo06}.

\begin{figure}[!b]
\centering\includegraphics[width=.8\columnwidth]{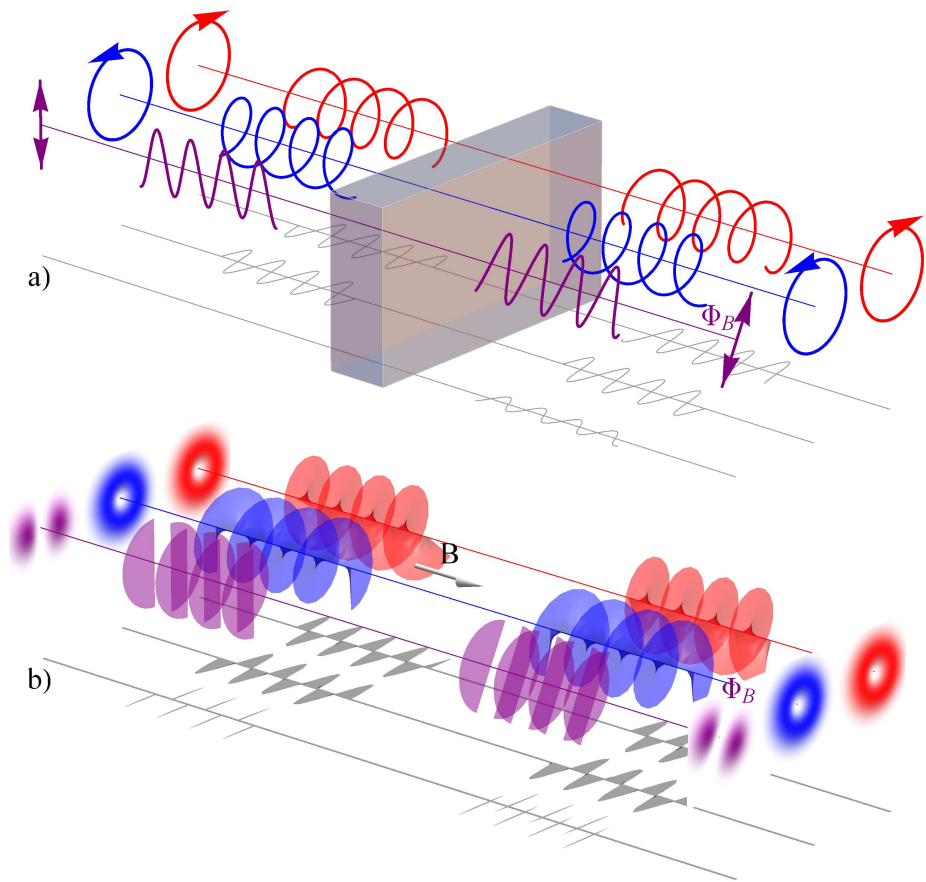}
\caption{\label{Faraday} (a) Illustration of the Faraday effect for spin angular momentum, {\it i.e.} optical polarization, where the two opposite
circular polarization components propagate at different speeds through an optically active medium in a magnetic field.  As a result, the linear
polarisation rotates by an angle proportional to the magnetic field.
(b) The analogous effect for transverse beam profiles of electrons, where the different propagation speed of states with opposite OAM in a longitudinal magnetic field leads to image rotation.}
\end{figure}

Faraday noted that the polarisation direction of light is rotated after passing through ``heavy glass" exposed to a longitudinal magnetic field.   We now understand that the Faraday effect arises from the different speed of propagation of right and left handed circularly polarised light through an optically active medium.  The associated difference in accumulated phase between the circular components of linearly polarised light results in a rotation of the polarisation direction. 

One of the intriguing properties of  light is that it can carry angular momentum: a spin contribution associated with circular polarisation  ($\pm \hbar$), but also orbital angular momentum (OAM)  \cite{Andrews2008, Franke-Arnold2008}.  While circular polarisation describes a rotation of the electric field vector upon propagation, the orbital angular momentum (OAM) is a feature of 'twisted' light beams. The OAM can take on  arbitrary multiples of $\hbar$ depending on how tightly wound the phase fronts are.  These so-called ``vortex beams" have a rotational intensity pattern and are associated with a phase dependence $\exp(il\phi)$, where $l$ is a non-zero integer and $\phi$ the azimuthal angle.

Strictly speaking,  Faraday rotation is not a relevant concept for optical OAM. The reason is that there is no {\it  intrinsic} mechanism in a
gyromagnetic medium to produce the required OAM state dependent dispersion, because selection rules forbid coupling of the OAM to the atomic electron
degrees of freedom. 
This is consistent with results from a recent experiment in which no rotation was observed for a superposition of right and left handed OAM states
(a Hermite-Gauss mode) propagating through cholesteric liquid crystals \cite{Loffler2011}. We note that a relative phase shift between  right and left handed OAM components  will appear as a rotation of the intensity pattern
\cite{Allen2007}.  Such phase shifts can be induced by spinning the medium through which the light propagates, inducing a `mechanical' Faraday rotation, as
demonstrated recently in a slow light medium \cite{Franke-Arnold2011}.   
 
\begin{figure}[!b]
\centering\includegraphics[width=.8\columnwidth]{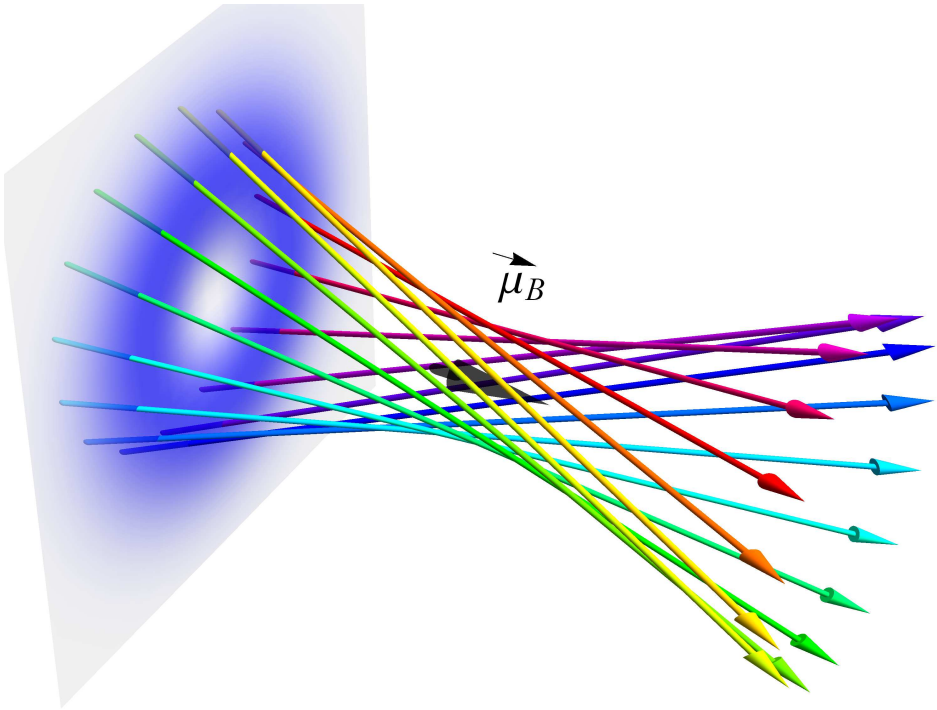}
\caption{\label{ElectronVortex} The circulating charge in an electron vortex beam generates a magnetic moment which interacts with external magnetic
fields.  The electron density function vanishes at the vortex core and the propagation direction is twisted around this core.}
\end{figure}

Electron vortices are unusual quantum states that have only recently been predicted \cite{Bliokh2007} and produced in transmission electron microscopy (TEM) experiments \cite{Uchida2010, Verbeeck2010}.  Electron vortex beams have the same geometrical properties as their optical counterparts, being characterised by an  $\exp(il\phi)$ angular dependence related to $l \hbar$ units of OAM, but they also produce features that have
no analogue in optics. In particular the circulation of charge
in an electron vortex beam gives rise to an arbitrarily large orbital magnetic moment (Fig. \ref{ElectronVortex}), distinct from the magnetic moment due to spin
\cite{Bliokh2007, PhysRevLett.107.174802}.  Hence electron vortices can couple to electronic degrees of freedom through dipole selection
rules forbidden to optical vortices \cite{Lloyd2012}. 

Given the analogies (and differences) between optical and electron vortices, the question arises: Do electron vortex waves undergo something analogous to
an optical Faraday effect? Here we show that there is indeed a Faraday rotation arising through Zeeman interaction from propagation \textit{parallel}
to a uniform, external magnetic field (i.e. in a geometry where there is no Lorentz force).

\section{Electron vortex states}

The dynamics of a non-relativistic electron propagating  in a uniform longitudinal magnetic field $\bm{B}$ (with associated vector potential $\bm A $)
is described by the  Hamiltonian 
\begin{equation}  %S1
\hat{H}=\frac{1}{2m}(-i\hbar\bm{\nabla}-e\bm{A})^2-\bm{B}\cdot\hat{\bm{\mu}}_S,
\end{equation}
where $m$ and $e=-|e|$ are the electron mass and charge, respectively.
The Hamiltonian contains the kinetic energy due to the canonical momentum and the Zeeman interaction of the electron spin 
with an arbitrary external magnetic field $\bm{B}$.  The electron's two-component spinor wavefunction $\tilde{\psi}$ satisfies the Pauli equation  $\hat{H}\tilde{\psi}
=i\hbar\frac{\partial}{\partial t}\tilde{\psi}$.
Here $\hat{\bm{\mu}}_S=
-g\mu_B\hat{\bm{\sigma}}/2$ is the operator for the magnetic moment,
$\mu_B=\hbar|e|/(2m)$ is the Bohr magneton, $g\approx 2$ is the Land\'{e} g-factor for electron spin
and $\hat{\bm{\sigma}}$ is the vector of Pauli spin matrices $(\hat{\sigma}_x,\hat{\sigma}_y,\hat{\sigma}_z)$.
In agreement with typical parameters in transmission electron microscopes ($\sim 8$ nA/nm$^2$), we assume that beam currents are sufficiently low so that Coulomb repulsion can be neglected. 

In the case of the uniform magnetic field directed along the $z$ axis, a suitable choice for the vector potential is
$\bm{A}=(B_zr/2)\hat{\bm{\phi}}$.  Exploiting the
cylindrical symmetry of the system, the Hamiltonian can be put in the form
\begin{equation}
\label{H}
\hat{H} = -\frac{\hbar^2}{2m}\frac{\partial^2}{\partial z^2}-\frac{\hbar^2}{2m}\nabla_{\perp}^2
+\frac{1}{2}m\omega_L^2r^2+\omega_L(\hat{L}_z+g\hat{S}_z). \label{Hcyl}
\end{equation}
Here $\omega_L=|e|B_z/(2m)$ denotes the Larmor frequency, $\hat{L}_z=-i\hbar\frac{\partial}{\partial\phi}$ and  $\hat{S}_z=s \hbar  \hat{\sigma}_z$ are the operators for the $z$ component of OAM and spin, respectively, where $s$ is the spin quantum number.
The first term gives the kinetic energy of motion along $z$, which is the same as in field-free space; the second and third terms together give the
energy for the transverse motion, and have the form of the Hamiltonian for a harmonic oscillator with characteristic frequency $\omega_L$; the final
term gives the Zeeman energy, with contributions from both OAM and spin.

With Hamiltonian (\ref{Hcyl}) the spinor components decouple and we can find monochromatic wave solutions which obey the time independent Schr\"{o}dinger equation  for a scalar wavefunction $\psi$.  
This problem can be solved exactly for eigenstates with  given $z$ components of OAM and
momentum \cite{landau1977}. Separating the degrees of freedom, 
\begin{equation} \label{psi}
\psi_{nls}(r,\phi,z)=R_{n|l|}(r)\exp(il\phi) \exp(ik_{nls}z)
\end{equation}
we identify the radial modes 
\begin{equation}
R_{n|l|}(r)=\sqrt{\frac{2n!}{\pi(n+|l|)!}}\frac{1}{w_B}\left(\frac{\sqrt{2}r}{w_B}\right)^{|l|}
{\rm e}^{-r^2/w_B^2} L_n^{|l|}\left(\frac{2r^2}{w_B^2}\right),
\end{equation}
where $w_B=2\sqrt{\hbar/|eB_z|}$ is a characteristic width which depends on the magnitude of the magnetic field, and $ L_n^{|l|}$ is an associated Laguerre polynominal.
The radial modes are characterised by the OAM quantum number, $l$, and the radial mode number $n=0,1,2,\ldots$ which denotes the number of
radial nodes of the electron density function.   The radial  profile of a mode with $l=1$ and $n=0$ is shown in Fig. 2.  
The transverse beam profile, given by $R_{n|l|}(r)\exp(il\phi)$, is the same as that of the Laguerre-Gauss beams familar from optical vortices, as was also pointed out in \cite{Bliokh2012}.

Recognising that the combined  second and third term in (\ref{Hcyl}) describe a 2D harmonic oscillator,
which in polar coordinates has energy eigenvalues $\hbar|\omega_L|(2n+|l|+1)$, the eigenvalues for the total energy are
 the well known Landau levels
\begin{equation} 
\label{E}
E=-\frac{\hbar^2k^2}{2m}+\hbar|\omega_L|(2n+|l|+1)+\hbar\omega_L(l+gs).
\end{equation}
The corresponding allowed wave numbers are then
\begin{equation}
\label{waven}
k_{nls}=k_0\sqrt{1-\frac{1}{E}[(2n+|l|+1)\hbar|\omega_L|+(l+gs)\hbar\omega_L]},
\end{equation}
where $E$ is the total energy determined by the electron source, and $ k_0=\sqrt{2mE}/\hbar$ is the wave number for a plane (non-vortex) wave propagating freely along the $z$ axis.
For electron vortices, the phase aquired upon propagation depends on the magnetic field. 
We see from (\ref{waven}) that there is a path that depends upon the direction of the angular momentum through the signs of $l$ and $s$. This
corresponds to a phase $\theta_{nls}=k_{nls}z$ acquired upon propagation which depends (via the Larmor frequency) on the magnetic field and is proportional to the total
angular momentum comprising both a spin and an orbital component.

If the magnetic energy is small compared to the total energy (a situation justified e.g. for electrons in TEM experiments), {\it i.e.} $(2n+|l|+1)\hbar|\omega_L|+(l+gs)\hbar\omega_L\ll E$, we can apply the paraxial approximation to the wave numbers (\ref{waven}).
%\begin{equation}  
%k_{nls}=k_0-(2n+|l|+1)|k_L|-(l+gs)k_L.
%\end{equation}
 The corresponding phase shift accumulated along the trajectory of the vortex then comprises three parts: 
\begin{equation}
\label{theta}
\theta_{nls}=k_0 z  -(2n+|l|+1)|k_{\rm L}| z-(l+gs)k_{\rm L} z,
\end{equation}
where $ k_L=\frac{m}{2E}\mu_B B_z/ \hbar$ is a spatial frequency which corresponds to the Larmor temporal frequency $\omega_L$. 
The first term describes the phase evolution in free space;  the second depends on the energy of the transverse motion due to the magnetic field, and the third term arises from the Zeeman interaction
with the total angular momentum.  In the following it is this latter term which is important, as it causes a different phase shift
for vortex states with opposite helicity.  We note that electrons in superpositions of left and right spin components should result in a spin Faraday rotation, analgous to optics.  Here  we concentrate on OAM Faraday rotation for which there is no optical counterpart.

Just like for the optical Faraday effect, the differential phase shifts become observable as a rotation angle for electrons in superpositions of vortex states with opposite handedness.  As they originate from the interaction of the magnetic dipole moment with the external field, the electron Faraday effect does not require the mediation of an optically active medium but occurs in vacuum!

\section{Considerations on observing the Faraday effect for electrons}

Electron vortex states have recently been generated in transmission electron microscopes (TEM) via diffraction from  nano-fabricated holograms
\cite{Verbeeck2010, McMorran2011,Verbeeck2012, Saitoh2012}.  Using suitably designed holograms, also superpositions of vortex states can be
generated  (see \ref{app grating}).  The required shape of the holograms is determined by the interference pattern of the target state
with a reference wave function, e.g. a plane or spherical wave, resulting in transverse or longitudinal separation of the diffraction orders
respectively. 

In order to realise electron Faraday rotation we require electrons in a superposition of two modes with the same
spin and radial mode number but opposite vorticity $\pm l$.  The probability density then has an azimuthal dependence
\begin{equation}
\frac{1}{2}\left| \psi_{nls}+\psi_{n(-l)s}\right|^2\propto\cos^2[l(\phi- \Phi_B)],
\end{equation}
where we define
\begin{equation}
\label{Phi}
\Phi_{\rm B}=k_{\rm L }z=\frac{1}{\hbar}\sqrt{\frac{m}{2E}}\mu_B B_z z.
\end{equation}

\begin{figure}[!t]
\centering\includegraphics[width=.8\columnwidth]{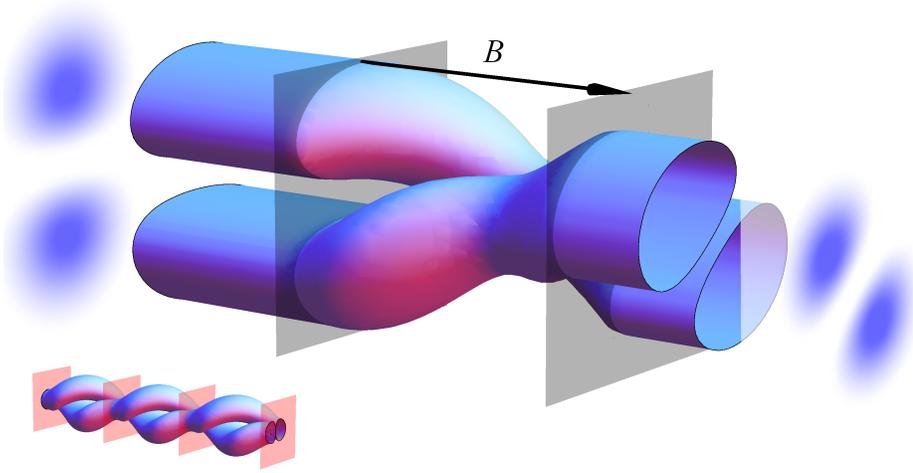}
\caption{\label{Faraday3D}The electron density distribution rotates when propagating through a parallel magnetic field, and at
the same time varies its beam waist periodically, with twice the period of the rotation (see inset).}
\end{figure}

For $n=0$ this is a petal pattern consisting of $2|l|$ maxima equally spaced around a circle, which after propagating through a region of a
longitudinal magnetic field is rotated through the angle $\Phi_{\rm B}$.
The maxima are separated by phase singularity lines, where the phase changes by $\pi$ and the probability density vanishes. 
For the  case of $l=\pm 1$ the transverse profile, shown in Fig.~\ref{Faraday3D}, is that of the ${\rm HG}_{10}$
Hermite-Gaussian mode.
%, and a three dimensional wavepacket would have the form of a $2p$ atomic orbital. 
Here the analogy with optical polarization is  clearest, with the $l=\pm 1$ components corresponding to the right and left handed circular
polarization states, and the nodal line to the linear polarization.

While the phase change depends on $l$, the rotation of the intensity pattern
is independent of $l$.  The Laguerre-Gauss modes form a complete basis with which an arbitrary wavefunction can be described,
and therefore any intensity and phase profile will rotate through the same angle $\Phi_{\rm B}$ \cite{Allen2007}.

So far we have considered the eigenstates of (\ref{H}) with a transverse scale determined by the magnetic field strength through
the parameter $w_B$.  A beam with the same radial profile but a width $w\neq w_B$  is, however, no longer an eigenstate and will therefore
change upon propagation.  Solving the paraxial wave equation, (see \ref{app paraxial}) we find that the radial profile retains the same Laguerre-Gauss form, only now expands and contracts periodically in time, or equivalently with propagation distance (\ref{waist}). This contraction  happens  at twice the Larmor frequency and hence  twice every full  rotation.  
 The width variation can be understood in terms of the competition between diffraction, which dominates when
$w<w_B$, and the confining effect of the magnetic field, which dominates when $w>w_B$.  This is illustrated in Figure 3.  

\begin{figure}[!b]
\begin{center}
\includegraphics[width=7cm,keepaspectratio=true]{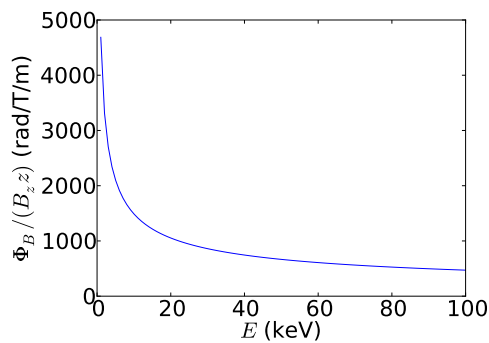}
\end{center}
\caption{Verdet parameter as a function of energy, i.e. the rotation angle per T and m.}
\label{Phi_vs_E}
\end{figure}

The rotation angle $\Phi_{\rm B}$, depends on the initial kinetic energy $E$ of the electrons, as detailed in (\ref{Phi}); as slow electrons
spend more time in the magnetic field, their rotation angle is larger.  While optical Faraday rotation is characterised by the Verdet constant (a proportionality constant of rotation angle
per propagation distance and magnetic field strength), the electron `Verdet' parameter varies with kinetic energy (see Fig.  \ref{Phi_vs_E}). 

Even with low-energy transmission electron microscopy, measuring the proposed deflection due to an interaction with a perpendicularly-magnetised sample remains challenging, in particular the neccessity to distinguish Faraday rotation from the usual cycloid motion of an electron beam within magnetic lenses. Ignoring relativistic corrections, 
typical values ($E=60$keV, sample thickness = $100$nm, longitudinal field $B = 1$T) yields a rotation about the beam axis of 0.06mrad. In certain geometries 
differential phase contrast techniques routinely measure deflections of this magnitude, when the rotation can be projected by long camera lengths to give measurable deflections. A more promising experimental approach will be to consider low energy electron beams. Low energy photoelectrons ejected by circularly polarised 
photons are also known to carry orbital angular momentum \cite{MacLaren09} and the subsequent propagation of these electrons through 
magnetic thin films may also visualise Faraday effects.

\section{Conclusions}

We have demonstrated that for electron vortex states propagating in a longitudinal magnetic field
the Zeeman interaction produces an OAM dependent dispersion. This results in a rotation of the probability density of a superposition of vortex states about the beam axis, analogous to
the optical Faraday effect. To the best of our knowledge this is a new concept, it is an effect that is not present in optics, and it may lead to applications in electron spectroscopy.  
 The magnitude of the rotation scales with the magnitude of the
magnetic field, and increases with decreasing energy. 
There are a number of interesting applications which may follow. OAM Faraday rotation provides the possibility of spatially
resolved measurements of longitudinal magnetic field components, analogous to the measurement of transverse fields in Lorentz microscopy,  by
measuring the rotation angle of a vortex superposition.
Moreover, we note that in the approximation considered here, spin and OAM are separately conserved. This would not be
expected for relativistic non-paraxial beams \cite{PhysRevLett.107.174802} or spatially varying magnetic fields
\cite{Gallup2001}, suggesting a route to investigating intrinsic spin-orbit coupling in an electron vortex beam,
via a spin-dependence of the rotation angle.

%*********Still to cite: \cite{Bliokh2012}

\ack
This work was supported by the Future and Emerging Technologies (FET)
program for Research of the European Commission, under the FET Open
grant agreement PHORBITECH No.~FP7-ICT-255914, and EPSRC Bridging the Gap. C.G. is supported under a SUPA Prize Studentship.
We thank D. MacLaren, D. McGrouther and S. McVitie for stimulating discussions.

\appendix

\section{Calculation of beam width variation using the paraxial approximation}
\label{app paraxial}
\setcounter{section}{1}

To find the propagation of a beam with a given beam waist we evaluate the time-independent Schr\"{o}dinger equation with Hamiltonian (\ref{H}) as before in the paraxial approximation.  We assume a solution of the form
\begin{equation} %S7
\label{psi_parax}
\psi(r,\phi,z)=u(r,\phi,z){\rm e}^{ik_0z},
\end{equation}
where $u(r,\phi,z)$ is an envolope function which describes the evolution of the beam profile upon propagation.
$k_0$ is as defined in the main text.  If $u$ varies sufficiently slowly with $z$  we can use the paraxial approximation
\begin{equation} %S8
\left|\frac{\partial ^2u}{\partial z^2}\right|\ll\left|2k_0\frac{\partial u}{\partial z}\right|,\left|\nabla_{\perp}^2u\right|.
\end{equation}
Then, substituting (\ref{psi_parax}) into the Schr\"{o}dinger equation, we arrive at the paraxial wave equation
\begin{equation}  %S9
\nabla_{\perp}^2u+2ik_0\frac{\partial u}{\partial z}-k_0^2k_L^2r^2u-
\frac{2k_0k_L}{\hbar}\hat{L}_zu=0.
\end{equation}
The first two terms here are the same as in the paraxial equation for an optical beam propagating in vacuum, or for an electron beam in field-free space.
The third term represents the confining effect of the magnetic field, and the final term gives the Zeeman interaction.
For an OAM eigenstate with $u\propto{\rm e}^{il\phi}$, the last term takes a constant value $-2k_0k_Llu$.
We can then factor out the phase due to the Zeeman interaction by writing
\begin{equation}  %S10
u=v\mbox{e}^{-ilk_Lz},
\end{equation}
for some function $v(r,\phi,z)$.  $v$ then satisfies the equation
\begin{equation}
\label{v}
\nabla_{\perp}^2v+2ik_0\frac{\partial v}{\partial z}-k_0^2k_L^2r^2v=0.
\end{equation}
This equation has the same form as the equation from paraxial optics for a medium with
quadratically varying refractive index $n(r)=n_0-\frac{1}{2}n_2 r^2$, where $n_0$ and $n_2$ are constants (see for example \cite{siegman86}),
only here $n_2\rightarrow k_L^2$.
Such an optical system supports Laguerre-Gauss type modes which experience a periodic width variation due to the competition between diffraction
and the focusing effect produced by the refractive index variation.  A similar effect is described in \cite{Gallatin2012}, for the
propagation of vortex wavepackets in a transverse field.

The solutions can be written \cite{Tien1965, Newstein1987} as
\begin{eqnarray}  %S11
\label{LG}
v_{nl}(r,\phi,z) & = & \sqrt{\frac{2n!}{\pi(n+|l|)!}}\frac{1}{w(z)}\left(\frac{\sqrt{2}r}{w(z)}\right)^{|l|}{\rm e}^{-r^2/w^2(z)} \nonumber \\
&  & L_n^{|l|}\left(\frac{2r^2}{w^2(z)}\right){\rm e}^{il\phi}\exp\left[-i\frac{k_0r^2}{2R(z)}\right] \nonumber \\
&  & {\rm e}^{-i(2n+|l|+1) \xi(z)},
\end{eqnarray}
where $w(z)$ is the beam width, $R(z)$ is the wavefront radius of curvature and $\xi(z)$ gives the longitudinal phase shift.  The equation (\ref{LG})
is the same as for the LG modes in free space, except here the functions $w(z)$, $R(z)$ and $\xi(z)$ are different.
Choosing $z=0$ to coinside with one of the minima of $w(z)$, and calling this minimum value $w_0$, the width function can be written, for the electron beam in
a magnetic field, as
\begin{equation}  %S12
\label{waist}
w(z)=w_B\sqrt{1-\left[1-\left(\frac{w_0}{w_B}\right)^2\right]\cos \left(2k_Lz\right)}.
\end{equation}

\section{Diffraction grating patterns for production of electron vortex superpositions}
\label{app grating}

\begin{figure}
\begin{center}
\subfigure[]{\label{linear_grating}\includegraphics[width=3.8cm]{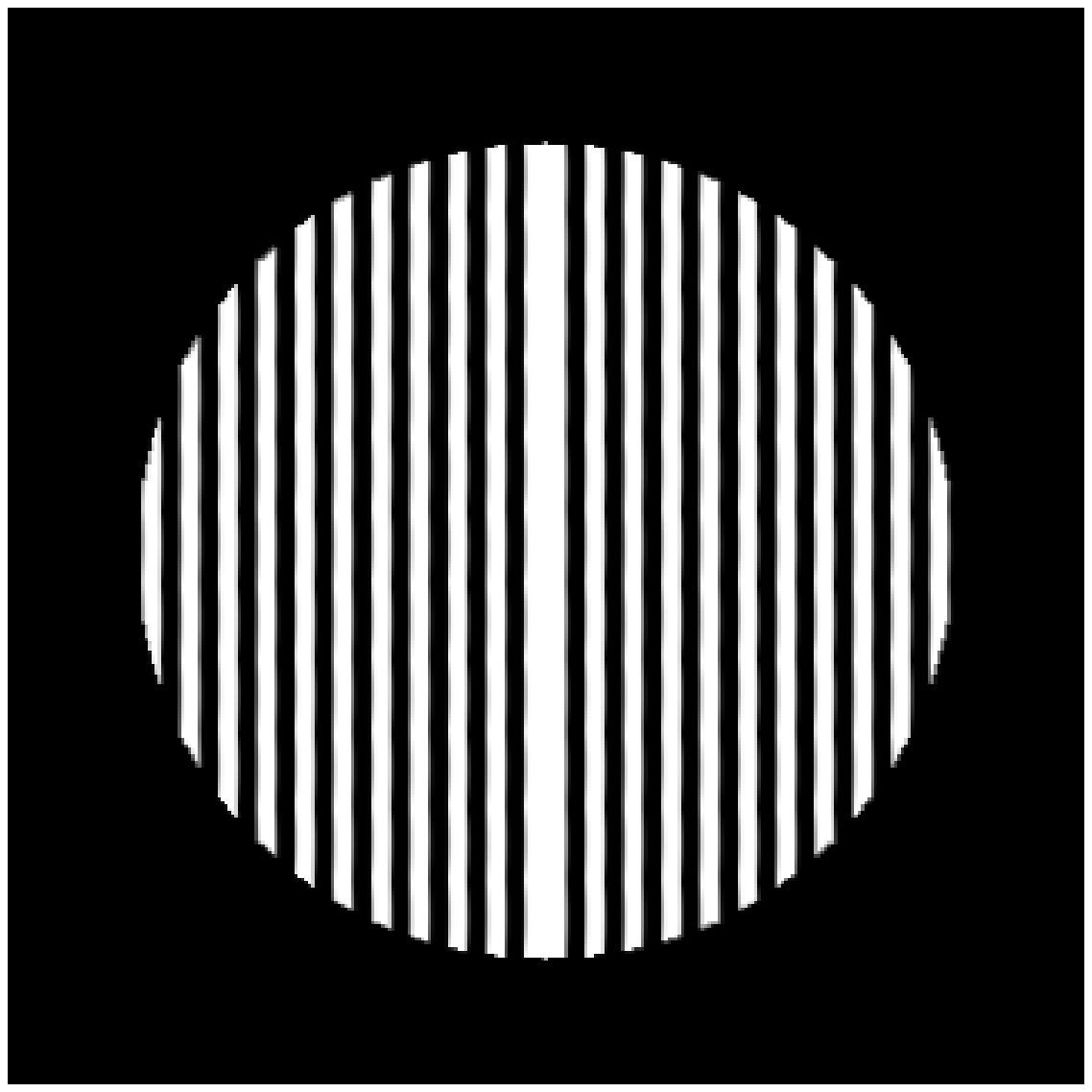}}
\subfigure[]{\label{spherical_grating}\includegraphics[width=3.8cm]{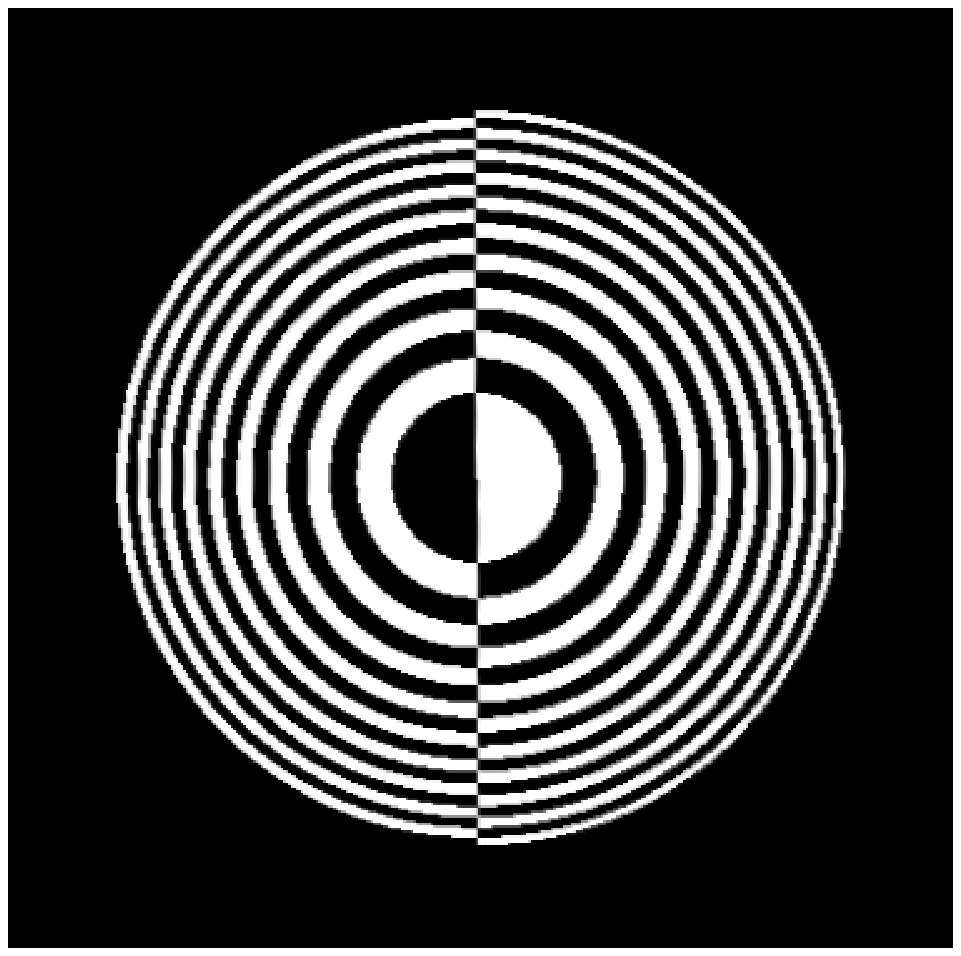}}
\subfigure[]{\label{diff_pattern}\includegraphics[width=8cm]{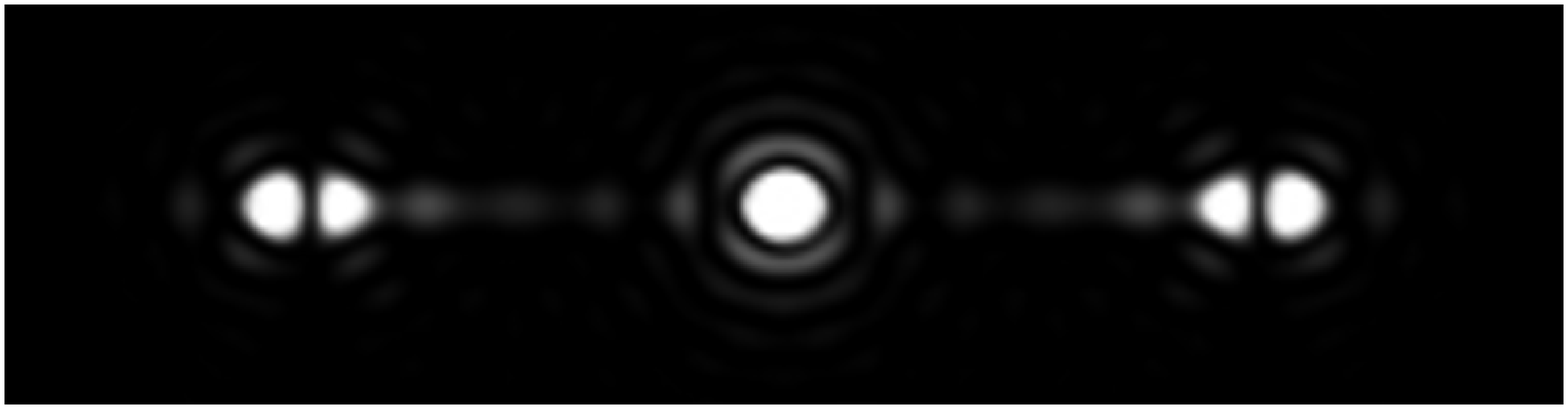}}
\end{center}
\caption{(a) Hologram grating to produce a superposition of vortex modes with $l=\pm1$. (b)
Grating to produce the same superposition but with the diffraction orders separated longitudinally along the beam
axis. (c) Diffraction pattern produced by grating in (a), with the desired superpositon in the $\pm1$ diffraction orders.}
\label{superpositions}
\end{figure}

For interference with a plane wave $\psi\propto{\rm e}^{ik_xx}$, in which case the diffraction orders will be separated
transversely, a suitable hologram pattern is generated from
\begin{equation}
\label{holo_eq}
|\psi^2|_{\rm holo} = \left\{ 
\begin{array}{rl}
 1 & {\rm if} \frac{1}{3}|2\cos l(\phi-\phi_0)+{\rm e}^{ik_xx}|^2 > \frac{1}{2} \\
 0 & {\rm otherwise},
\end{array} \right.
\end{equation}
where $\phi_0$ specifies the orientation of the singularity lines.  A grating producing a superposition of $l=1$ and $l=-1$
modes in the first diffraction order is shown in
Fig.~\ref{linear_grating}, where we have chosen $\phi_0=0$.  Note that the left and right sides of the grating are displaced
by half a period with respect to one
another, introducing the necessary $\pi$ phase shift.
In a similar way a hologram which separates the diffracted beam components longitudinally rather than transversely could
be used \cite{Verbeeck2012, Saitoh2012}.  An example of this type of grating is shown in Fig.~\ref{spherical_grating}.
This is calculated using the same method, only with the plane wave factor $\psi\propto{\rm e}^{ik_xx}$
replaced with the spherical wave profile $\psi\propto{\rm e}^{iCr^2}$, where $C$ is a constant which determines the curvature
of the wavefront, and hence the spacing of the diffraction orders.
% Note that for gratings which separate the diffraction orders longitudinally, the interference between different
% orders will not affect the position of the singularity lines.
Such grating patterns can be produced in the same way as those
already used to generate single vortex modes.  We note that gratings that generate superpositions of vortex beams with beams that have flat phasefronts have recently been generated and used to investigate the Gouy phase for electrons \cite{MacMorranFiO}. 

\vspace{5mm}

\bibliographystyle{unsrt} % Bibliography style file, unsrt.bst
\bibliography{ElectronFaraday}

\end{document}